\documentclass[%
reprint,
 amsmath,amssymb,
 aps,
 prb,
]{revtex4-1}

\usepackage{graphicx}
\usepackage{dcolumn}
\usepackage{bm}
\usepackage{xcolor}
\usepackage[unicode=true,colorlinks=true,linkcolor=blue,citecolor=blue, urlcolor= blue]{hyperref}

\begin{document}

\preprint{APS/123-QED}

\title{Atomic scale modeling of the thermodynamic and kinetic properties \\of dilute alloys driven by forced atomic relocations}

\author{Liangzhao Huang}
 \email{liangzhao.huang@cea.fr}
\author{Thomas Schuler}
\author{Maylise Nastar}
\affiliation{%
 CEA, DEN, Service de Recherches de M\'etallurgie Physique, Universit\'e Paris-Saclay, 91191, Gif-sur-Yvette, France
}%

\date{\today}

\begin{abstract}
{Sustained external forces acting on a material provide additional mechanisms to evolve the state of the system, and these mechanisms do not necessarily obey the microscopic detailed balance. Therefore, standard methods to compute the thermodynamic and kinetic properties do not apply in such driven systems.} The competition between these mechanisms and the standard thermally activated jumps leads to non-equilibrium steady states. {We extend the Self-Consistent Mean Field theory to take into account forced atomic relocations (FARs) as a model of these additional kinetic mechanisms.} The theory is applied to the atomic-scale modelling of radiation damage. {Using a first-shell approximation of the theory, we highlight the violation of Onsager reciprocal relations in driven systems.} An implementation of the extended theory into the KineCluE code yields calculations of the effective Onsager coefficients in agreement with Monte Carlo simulations. {A systematic parametric study is performed to emphasize the effect of FAR distances and the solute-defect interaction on the diffusion properties.} The effect of FAR on the vacancy-solute flux coupling and the solute tracer diffusivity is non-negligible when: (i) the solute-vacancy thermodynamic attraction is large, (ii) {the magnitude of the thermal jump frequencies is lower or comparable to the frequencies of FAR}, and (iii) the range of interactions between vacancies and solute atoms is close to FAR distances.

\end{abstract}

\maketitle


\section{Introduction} \label{sec:introduction}

Excess point defects (PDs) are massively generated in materials driven away from equilibrium by external forces such as mechanical solicitation and irradiation~\cite{Martin1996,Gary2007}. This phenomenon induces fluxes of PDs toward the microstructural features (e.g., grain boundaries, dislocation lines, etc.) acting as PD sinks. The solute-PD interaction in alloys leads to the coupling between the PD and solute atom fluxes to the sinks. This flux coupling is the main kinetic process controlling the redistribution of solute atoms in alloys driven by an excess of PDs~\cite{Anthony1968,Anthony1969,Okamoto1974,Barbu1975,Okamoto1979,Kato1992,Bruemmer1999,Nastar2012,Ardell2016}. 

{Additionally, the external forces may introduce new mechanisms that affect the state of the system at the atomic scale. Examples of these mechanisms include collective motion of atoms or mixing. For instance, the latter mechanisms occur under irradiation producing displacement cascades~\cite{Haff1977,Motta1992,Averback1998} or during severe plastic deformation such as shearing~\cite{Vo2013,Ashkenazy2017}, torsion~\cite{Pouryazdan2012,Beach2017} and ball milling~\cite{Pochet1995,Klassen1997,Suryanarayana2004}. Unlike the thermally activated mechanisms leading the system toward equilibrium, these additional mechanisms are mostly athermal. They do not obey the microscopic detailed balance and lead to the disordering of the atomic configurations~\cite{Martin1996}. Note that both the thermally activated and externally forced mechanisms contribute to the mass transport in driven alloys, such that the system is driven to a non-equilibrium steady state (NESS)~\cite{Martin1996}. This situation prevents the use of standard methods to compute the thermodynamic and kinetic properties. In order to understand and predict the effect of external driving forces, it is crucial to be able to model the interplay between the two above-mentioned mechanisms. However, there is no model of flux coupling simultaneously accounting for these two competitive mechanisms at the atomic scale.}

{Radiation damage is a good school case for the modeling of driven alloy thermodynamics and kinetics because there exists microscopic models that effectively reproduce the atomic mixing caused by the external force~\cite{Martin1984,Averback1986}. Under irradiation, atoms are regularly hit by incident particles. The collision transfers kinetic energy from the incident particle to the primary knock-on atom (PKA). If this energy is below the displacement threshold energy (DTE), the PKA will only vibrate around its position unless a PD is located nearby, in which case the atom may exchange its position with the PD~\cite{Roussel2002}. For a recoil energy well above DTE (e.g., typically 1\,keV in metals), the PKA will move away from its original site, thereby creating Frenkel pairs and transferring kinetic energy to neighbouring atoms, which will themselves move away from their positions, so on so forth. Locally, this displacement cascade process produces a large excess number of PDs as well as atomic mixing since most of the atoms are displaced~\cite{Littmark1980}. The climax of this process is called the heat spike, where the material is locally liquid-like and atoms are able to rapidly diffuse in this region~\cite{Vineyard1976}. The excess energy eventually dissipates leading to a quench-like process where the crystalline structure is recovered and only a small excess number of PDs remain~\cite{Brinkman1954, Benedek1987, Nordlund1999}. The whole phenomenon can be effectively modeled by (1) creating PDs and PD clusters, and (2) shuffling atomic positions. The latter process is often considered random~\cite{English1981,Pramanik1986} but there are some evidences that it is in fact affected by thermodynamics~\cite{Workman1987}. Besides, recent investigations on concentrated alloys~\cite{Terentyev2006,Aidhy2015,Zhang2017} and high entropy alloys~\cite{Do2018} have shown that during the quenching stage, the spatial distribution of the PDs compared with the solute atoms is partially driven by the thermodynamic short range order (i.e. binding interaction). For more details concerning displacement cascades, please refer to Ref.\,\cite{Nordlund2018}.}

{The mixing of atomic positions in the displacement cascade was previously modeled effectively by forced atomic relocations (FARs) which consist in forced exchanges of positions between an atom and its nearest neighbour species such as another atom or a PD~\cite{Martin1984,Enrique2000,Lear2017,Soisson2018}. However, details in the cascade such as the spatial correlation between solute atoms and PDs due to the thermodynamic interactions were neglected. Moreover, the thermal mechanism and FAR were considered separately at the macroscopic scale. For instance, the tracer diffusion coefficient of the solute atom was written as the sum of two diffusion coefficients respectively related to the two mechanisms~\cite{Gary2007}. In this case, the interplay between these two mechanisms was neglected. An improved version of the model was proposed in Ref.\,\cite{Roussel2002}, where the five-frequency model~\cite{Lidiard1955,Lidiard1960} in face-centered cubic (fcc) systems was generalized to account for both types of kinetic mechanisms simultaneously. However, solute diffusivity is the only accessible quantity from this model, so it does not provide any information on solute-PD flux coupling. Moreover, long range FAR cannot be accounted for since the five-frequency model is limited to first nearest neighbour (1NN) interactions.} 

Thermodynamic and kinetic properties such as flux coupling coefficients and tracer diffusion coefficients of an alloy can be calculated from the Onsager matrix of the transport coefficients. Whenever the diffusion mechanism satisfies the microscopic detailed balance, Onsager has demonstrated that this matrix is symmetric~\cite{Onsager1931-1,Onsager1931-2}. We may calculate it either from the equilibrium atomic displacement fluctuations using the Allnatt formulae~\cite{Allnatt1965,Allnatt1982} or from the flow of matter resulting from an applied external force. However, for driven alloys including athermal mechanisms not obeying the microscopic detailed balance, we cannot compute the transport coefficients by means of a Monte Carlo numerical approach based on the Allnatt formulae. Yet, recent statistical theories have shown that it is possible to derive an effective Onsager matrix from the fluctuation theorem~\cite{Evans1993,Gallavotti1996}, though the resulting matrix is non symmetric. These theories go beyond the linear response theory. They provide a methodology for the investigation of far from equilibrium kinetics. Such an approach has been applied to the study of a molecular motor driven by forced chemical reactions~\cite{Lau2007,Lacoste2008}. However, it is not directly applicable to properly model systems with FAR because there are no notions of alloying effects and kinetic correlations in this model. In the context of research on diffusion in alloys, one knows how to deal with the complexity of calculating a sequence of PD jumps when the frequency of each jump depends on the local environment of the defect as long as the diffusion mechanism satisfies the microscopic detailed balance~\cite{Allnatt1993,Nastar2005,Garnier2013,Garnier2014,Messina2014,Messina2016,Schuler2016,Abhinav2019}. 

To obtain the transport coefficients at NESS, we start from the standard self-consistent mean field (SCMF) theory~\cite{Nastar2000,Nastar2005}, which was applied to the calculation of transport coefficients only in systems near equilibrium with jump mechanisms obeying the microscopic detailed balance. {In this paper, we generalize the SCMF theory by including the athermal mechanism of FAR. The generalized SCMF is implemented in the KineCluE code~\cite{Schuler2020} in order to perform automated calculations of transport coefficients. Different models of FAR are tested and the impact of each model parameter is systematically investigated. This study allows us to understand the conditions in which FAR significantly affect the material thermodynamic and kinetic properties.} 

The structure of the paper is as follow. In Sec.\,\ref{Sec_Models} we introduce the model of irradiation damage and thermal diffusion. Then, we introduce the mean field kinetic model to estimate the PD concentration under irradiation. Sec.\,\ref{sec_SCMF} is devoted to the calculation of the transport coefficients within the SCMF framework. In Sec.\,\ref{sec:results}, we focus on the comparison between the results given by different FAR models in various representative model alloys. The impact of different model parameters are studied in Sec.\,\ref{sec:sensitivity_relocation_model}. A discussion of the results, including the limitations and possible improvements, is presented in Sec.\,\ref{Sec_Discussions}.  

\section{Modeling of diffusion mechanisms under irradiation} \label{Sec_Models}

\subsection{Thermally activated jump frequencies}\label{subsec:thermal_jump}
We use the transition state theory to model thermally activated diffusion~\cite{Vineyard1957}. We introduce the thermal jump frequency $\omega_{\textbf{n}\rightarrow \widetilde{\textbf{n}}}^{\alpha V}$ associated with the thermally activated exchange of atom $\alpha$ and vacancy V which brings the system from configuration $\textbf{n}$ to $\widetilde{\textbf{n}}$:
\begin{equation}
\omega_{\textbf{n}\rightarrow \widetilde{\textbf{n}}}^{\alpha V} = \nu\,\text{exp}\left(-\frac{E^{\text{mig}}_{\textbf{n}\rightarrow \widetilde{\textbf{n}}}}{k_B T}\right),
\end{equation} where $\nu$ is the attempt frequency, $k_B$ is the Boltzmann constant, $T$ is the temperature and $E^{\text{mig}}_{\textbf{n}\rightarrow \widetilde{\textbf{n}}}$ is the migration barrier from configuration $\textbf{n}$ to $\widetilde{\textbf{n}}$, which can be computed by means of ab initio calculations~\cite{Tucker2010,Messina2014}. This mechanism is mediated by PDs and the jump rate depends on the temperature as well as the initial and saddle-point configurations. Note that it satisfies the principle of the microscopic detailed balance:
\begin{equation}
    P_{\textbf{n}}\,\omega_{\textbf{n}\rightarrow \widetilde{\textbf{n}}}^{\alpha V} = P_{\widetilde{\textbf{n}}}\,\omega_{\widetilde{\textbf{n}}\rightarrow \textbf{n}}^{\alpha V},
\end{equation}
where $P_{\textbf{n}}$ is the probability of configuration $\textbf{n}$. 

{
\subsection{Irradiation damage} \label{subsec:cascade_models}
We follow the ideas of previous studies to model the radiation damage by FAR. The latter includes two mechanisms: (1) FAR between two randomly chosen atoms (FAR-a) which consists in exchanging the positions of two atoms on lattice sites, and (2) FAR between a randomly chosen atom and a PD (FAR-d) which consists in exchanging the positions between an atom and a vacancy (V) or a self-interstitial atom (SIA). We need to account for the removing and creation of PDs within a cascade. {Here we consider that the PDs ``disappear'' during the heat spike, because the material is locally a liquid-like phase where there is no notion of PD. Later, during the process of quenching, only a small fraction of the PDs ``reappears'' somewhere in the cascade area, as if they had effectively jumped to another crystalline site during the cascade. The effective result of this process is modeled by FAR-d.}

For recoil energy well above DTE producing displacement cascade, the overall effect of the mixing is modeled by FAR characterized by a relocation distance $r$. FAR occurs at a given frequency proportional to the radiation flux.} First for FAR-a, we assume that the probability density function $p(r)$ of the relocation distance follows an exponential decay~\cite{Enrique2000,Enrique2003,Demange2017}:
\begin{equation}
    p\,(r) = \frac{1}{r_\text{m}}\,\text{exp}\left(-\frac{r}{r_\text{m}}\right),
\end{equation}
where $r_\text{m}$ is the mean relocation distance which is related to the size of the displacement cascade. Note that the latter depends on the material and the recoil energy of PKA. For example, the sizes of cascades generated in metals by fast neutrons or by heavy ions typically range between 10 and 100\,\r{A}~\cite{English1981,Phythian1995}. At the atomic scale, the relocation distance $r$ is discrete and is equal to one of the $i$-th NN distances. We define the probability mass function $\mathcal{P}(i)$ so that the distribution $p(r)$ in the interval $[r_{i},r_{i+1}]$ is averaged to the $i$-NN point:
\begin{equation}
    \mathcal{P}(i)=\int_{r_i}^{r_{i+1}}p\,(r)\,\text{d}r,
\end{equation}
where $r_i$ corresponds to the $i$-NN distance. In practice, we consider only a finite set of nearest neighbours, meaning that there is a cut-off relocation distance $L$-NN beyond which the probability is set to 0. In this case, we define the normalized probability mass function $\mathcal{P}_L(i)$ as:
\begin{equation}\label{eq:model_2_distribution}
    {\mathcal{P}}_L(i) = \frac{{\mathcal{P}}(i)}{\sum_{s=0}^{L}{\mathcal{P}}(s)}.
\end{equation}
We introduce as well a simplified model associated with a single relocation distance $r_m$ because it gives access to an analytical solution. 

{Here we ignore FAR-d of SIA and this assumption is justified in Sec.\,\ref{subsec:forced_relocation}. Therefore, we consider only FAR-d of vacancy. We propose two categories of FAR-d models: either the same relocation model employed for FAR-a, or a model favoring the relocation sites close to the solute atoms in case of attractive binding energies between vacancy and solute atoms. The latter model makes sense because in the quench-like process at the end of the displacement cascade, the remaining PDs form preferentially where their formation energy is the lowest, that is in the vicinity of solute atoms.} 

In order to represent both categories of models, we introduce three models. Models 1 and 2 for the first category, and Model 3 for the second category including a thermodynamic effect on FAR-d. Model 1 includes a single relocation distance for both solute and vacancy, while Model 2 includes an exponential law for the relocation distance (Eq.\,\eqref{eq:model_2_distribution}) for both species. {Model 3 is similar to Model 2, the only difference is that when the relocated vacancy is located at a distance lower than a threshold value $R_c$ from the solute atom B, the vacancy is systematically exchanged with an atom randomly chosen among the 1NN atoms of B (chemically biased FAR-d).

For recoil energy below DTE, the effective result of the sub-threshold collision is modeled only by FAR-d. The model of FAR-d is the same in Model 1 while FAR-a is not performed. The relocation distance is set to 1NN distance $r_1$. } 

\subsection{FAR frequencies}\label{subsec:forced_relocation}
{The FAR-d frequency is denoted $\Gamma^\text{ad}$, and the FAR-a frequency is denoted $\Gamma^\text{aa}$.  

When the recoil energy is above DTE, the relocation frequency $\Gamma^{aa}$ can be deduced from the radiation dose rate $\phi$ based on the ion-beam mixing framework~\cite{Haff1977,Averback1986}. In our model, FAR-a reproduces the mixing of atoms in the displacement cascade, which is related to the number of PDs produced by the PKA. After the quenching phase, there is only a small fraction of surviving defects, which defines the unit of displacement per atom (dpa). Therefore, there is a factor $n_\text{rel}$ relating $\Gamma^\text{aa}$ and the radiation dose rate in unit of dpa/s (see Eq.\,\eqref{eq:n_rel}). 
\begin{equation} \label{eq:n_rel}
    \Gamma^\text{aa}=n_\text{rel}\,\phi.
\end{equation}
From the literature, we set $n_\text{rel}= 100$~\cite{Riviere1983,Averback1986,Muller1988}. The latter number varies with the alloy thermodynamics due to the thermal effect on the atomic mixing rate in the cascade~\cite{Workman1987}. The frequencies of FAR-a and FAR-d depend on the number of cascades formed per unit of time as stated in Section\,\ref{subsec:cascade_models}. Therefore, $\Gamma^\text{aa}$ and $\Gamma^\text{ad}$ are both proportional to the dose rate. Hence, they are proportionally related by:
\begin{equation}
    \Gamma^\text{ad} = \gamma\,\Gamma^\text{aa},
\end{equation}
with $\gamma$ the proportionality constant. Note that $\gamma$ is set to 1 if not specified, i.e. $\Gamma^{ad}=\Gamma^{aa}=\Gamma$. Sensitivity studies concerning the value of $\gamma$ are shown in Sec.\,\ref{subsec:sensibility_PD_frequency}.

When the recoil energy is below DTE, the sub-threshold irradiation does not induce FAR-a because no displacement cascade is produced, thus $\Gamma^{aa}=0$. In this case, the calculation of $\Gamma^\text{ad}$ is not related to $\Gamma^{aa}$ and is directly deduced from the recoil energy. 
}

Note that the maximum dose rate under realistic irradiation condition is around 1\,dpa/s~\cite{Gary2007}, thereby leading to a relocation frequency of about 100\,s$^{-1}$. The latter is still very small compared to the thermal jump frequency of SIA, even at low temperature. For instance, the SIA thermal jump frequency in pure nickel at 300\,K is around $10^{10}$\,s$^{-1}$ according to the atomic diffusion data given in Ref.\,\cite{Tucker2010}. Therefore, we do not expect an important impact of FAR-d on the SIA-mediated diffusion properties. Therefore, we consider only FAR-d with vacancies, as stated in Sec.\,\ref{subsec:forced_relocation}. Yet, we emphasize that the extension of our framework to account for FAR-d of SIAs is straightforward.

\subsection{Point defect concentration}\label{subsec:PD_concentration}
The global concentration of PD varies under irradiation, mainly due to the production of Frenkel pairs, the mutual recombination between SIA and V, the elimination of PD at PD sinks such as grain boundaries and dislocations. The vacancy concentration at NESS $C_{V}^{\text{ness}}$ is estimated from a rate theory model~\cite{Russell1984,Schuler2017-2}: 
\begin{align} \label{eq:V-phi}
C_{V}^{\text{ness}}=C_{V}^{\text{eq}}-\dfrac{k^2\Omega}{8\pi r_c}+\sqrt{\left(\dfrac{k^2\Omega}{8\pi r_c} \right) ^2 + \dfrac{\phi\,\Omega}{4\pi r_c D_V} },
\end{align} 
where $C_{V}^{\text{eq}}$ is the thermal vacancy concentration at equilibrium, $r_c$ is the SIA-V recombination radius usually assumed to be of the order of the lattice parameter $a_0$. $\Omega$ is the atomic volume. $\phi$ is the radiation dose rate. $k^2$ is the sink strength assumed to be constant with the radiation dose rate with typical values ranging from $10^{12}$ to $10^{19}$~m$^{-2}$~\cite{Soisson2016} and $D_V$ is the vacancy diffusion coefficient. Note that the equilibrium concentration $C_{V}^{\text{eq}}$ is obtained from the vacancy formation enthalpy $H_V^\text{f}$ and entropy $S_V^\text{f}$ by:
\begin{equation}
    C_{V}^{\text{eq}}=\text{exp}\left( -\frac{H_V^\text{f}-T\,S_V^\text{f}}{k_B\,T} \right).
\end{equation}
We may then replace the flux $\phi$ by its expression in terms of $\Gamma$ from Eq.\,\eqref{eq:n_rel} into Eq.\,\eqref{eq:V-phi},
leading to a direct relationship between the vacancy concentration at NESS and $\Gamma$.

\section{Diffusion theory}\label{sec_SCMF}
For total exchange frequencies, we use the notation: 
\begin{equation} \label{eq:total_frequencies}
    W_{\textbf{n}\rightarrow \widetilde{\textbf{n}}}=W_{\textbf{n}\rightarrow \widetilde{\textbf{n}}}^{AV}+W_{\textbf{n}\rightarrow \widetilde{\textbf{n}}}^{BV}+W_{\textbf{n}\rightarrow \widetilde{\textbf{n}}}^{AB},
\end{equation}
where $W_{\textbf{n}\rightarrow \widetilde{\textbf{n}}}^{AV}=\omega_{\textbf{n}\rightarrow \widetilde{\textbf{n}}}^{AV}+\Gamma_{\textbf{n}\rightarrow \widetilde{\textbf{n}}}^{AV}$, $W_{\textbf{n}\rightarrow \widetilde{\textbf{n}}}^{BV}=\omega_{\textbf{n}\rightarrow \widetilde{\textbf{n}}}^{BV}+\Gamma_{\textbf{n}\rightarrow \widetilde{\textbf{n}}}^{BV}$ and $W_{\textbf{n}\rightarrow \widetilde{\textbf{n}}}^{AB}=\Gamma_{\textbf{n}\rightarrow \widetilde{\textbf{n}}}^{AB}$. Note that, although relocation frequencies do not depend on the configuration before and after the exchange, for the sake of clarity, we choose to follow the notation of the thermal jump frequencies. 

We start with a Master Equation expressing the fact that the probability distribution of different configurations is controlled by the transition probabilities between two configurations:
\begin{equation} \label{eq:master_equation}
\dfrac{\text{d} }{\text{d}t} \bm{P}=\bm{W} \bm{P},
\end{equation} where $\bm{W}$ is a matrix with components $\bm{W}_{\textbf{n}\widetilde{\textbf{n}}} = W_{\widetilde{\textbf{n}}\rightarrow \textbf{n}}$ if $\textbf{n}\neq \widetilde{\textbf{n}}$ and $\bm{W}_{\textbf{n}\textbf{n}}=-\sum_{\widetilde{\textbf{n}}\neq \textbf{n}}W_{\textbf{n}\rightarrow \widetilde{\textbf{n}}}$. $\bm{P}=(P_\textbf{n})$ is a linear vector of probabilities of configurations ($\textbf{n}$).

{For now, the recombination reactions between SIA and V are introduced at the upper scale, within the mean field rate theory model of the average PD concentrations (see Sec.\,\ref{subsec:PD_concentration}). These athermal events are not treated on the same foot as FAR because SIA and V are considered to be well-separated at the end of the displacement cascade under dilute approximation~\cite{Nordlund2018}. In this case, the SIA-V recombination requires long-range diffusion, thereby not incorporated in the microscopic Master Equation. }

We explain below the method we use to determine the dynamical short range order (SRO) parameters at NESS and the diffusion properties from the Master Equation.

\subsection{Dynamical short range order}\label{subsec:SRO}

Starting from the thermal equilibrium state, the mix of thermal jumps and FAR leads to NESS. The latter state is characterized by dynamical SRO parameters which depend on FAR frequencies and thermal jump frequencies. We define them from the configurational probabilities, deduced from a stationary condition applied to the Master Equation (Eq.\,\eqref{eq:master_equation}), also called the global detailed balance condition:
\begin{equation} \label{eq:global_detailed_balance}
    \forall \textbf{n},~ \sum_{\widetilde{\textbf{n}}}W_{\widetilde{\textbf{n}}\rightarrow \textbf{n}}P_{\widetilde{\textbf{n}}} - W_{\textbf{n}\rightarrow \widetilde{\textbf{n}}}P_{\textbf{n}} = 0.
\end{equation}
The solution of Eq.\,\eqref{eq:master_equation} at NESS is noted $\bm{P^\text{ness}}=(P_\textbf{n}^\text{ness})$. {The SRO parameter for configuration $\textbf{n}$ is defined as the ratio between the configurational probability $P_\textbf{n}^\text{ness}$ and the one of the a reference configuration denoted $P_{0}^\text{ness}$.}

\subsection{Transport coefficients} \label{subsec_Lij}
The phenomenological transport coefficients ($\lambda_{\alpha\beta}$) are fundamental parameters to describe the diffusion of chemical species ($\alpha$, $\beta$) in alloy at the macroscopic scale. Fluxes of chemical species ($J_\alpha$) are proportional to these coefficients:
\begin{equation}
    \overrightarrow{J}_\alpha=-\sum_\beta \lambda_{\alpha\beta}\frac{\overrightarrow{\nabla}\mu_\beta}{k_B T},
\end{equation}
with $\nabla\mu_\beta$ the established driving force deviating the system from equilibrium. Starting from NESS, we apply a small gradient of chemical potential and compute the resulting flux of atoms and vacancy. Here we extend the SCMF model to jump mechanisms not obeying the microscopic detailed balance. The SCMF theory was first proposed to study the diffusion process with atomic jumps following the principle of microscopic detailed balance~\cite{Nastar2005} but the latter is broken because of FAR. By following the nomenclature of Ref.\,\cite{Nastar2005}, the configuration is defined by a vector $\textbf{n}$. The latter consists in occupation numbers of all species on all sites i.e. \{$n^{\text{A}}_1$,$n^{\text{B}}_1$,$n^{\text{V}}_1$; $n^{\text{A}}_2$,$n^{\text{B}}_2$,$n^{\text{V}}_2$; ...\}, with $n^\alpha_i$ equal to one if the site $i$ is occupied by species $\alpha$ and zero if not. {The transition from configuration $\textbf{n}$ to $ \widetilde{\textbf{n}}$ is realized by thermally activated jumps or FAR, with a total frequency ${W}_{\textbf{n} \rightarrow \widetilde{\textbf{n}}}$.} Within the standard SCMF theory in Ref.\,\cite{Nastar2005}, $P_\textbf{n}(t)$, the non-equilibrium distribution function of configuration $\textbf{n}$, is expressed as the product of the equilibrium probability $P^\text{eq}_\textbf{n}$ and a non-equilibrium contribution. Here we choose the reference state to be NESS, and replace $P^\text{eq}_\textbf{n}$ by the probability distribution function $P^{\text{ness}}_\textbf{n}$:
\begin{equation}
    P_\textbf{n}(t) = P^{\text{ness}}_\textbf{n}\times\delta P_\textbf{n}(t).
\end{equation}
The Master Equation (see Eq.\,\eqref{eq:master_equation}) is written for a certain configuration $\textbf{n}$ as
\begin{equation} \label{eq:master_equation_microscopic}
    \frac{\text{d}P_\textbf{n}(t)}{\text{d}t}=\sum_{\widetilde{n}}\left[ {W}_{{\widetilde{\textbf{n}} \rightarrow \textbf{n}}} P^{\text{ness}}_{\widetilde{\textbf{n}}}\delta P_{\widetilde{\textbf{n}}}(t) - {W}_{\textbf{n} \rightarrow \widetilde{\textbf{n}}} P^{\text{ness}}_\textbf{n}\delta P_\textbf{n}(t) \right].
\end{equation}
By applying the global detailed balance condition, i.e. Eq.\,\eqref{eq:global_detailed_balance}, we obtain a reformulation of the Master Equation:
\begin{equation} \label{eq:master_equation_microscopic_2}
    \frac{\text{d}P_\textbf{n}(t)}{\text{d}t}=\sum_{\widetilde{n}} {W}_{{\widetilde{\textbf{n}} \rightarrow \textbf{n}}} P^{\text{ness}}_{\widetilde{\textbf{n}}}\left[ \delta P_{\widetilde{\textbf{n}}}(t) - \delta P_\textbf{n}(t) \right].
\end{equation}
Note that the standard SCMF theory relies on the microscopic detailed balance (${W}_{{\widetilde{\textbf{n}} \rightarrow \textbf{n}}} P^{\text{ness}}_{\widetilde{\textbf{n}}}={W}_{\textbf{n} \rightarrow \widetilde{\textbf{n}}} P^{\text{ness}}_\textbf{n}$). In that case, it is equivalent to consider the transition probabilities entering or exiting a given configuration. When the microscopic detailed balance is not satisfied the transition frequencies to be retained are the entering configurations. The derivation from the Master Equation (Eq.\,\eqref{eq:master_equation_microscopic_2}) of the transport coefficients is similar to the standard SCMF theory in Ref.\,\cite{Nastar2000,Nastar2005}. It is summarized in Appendix. 

\subsection{SCMF theory under first shell approximation}\label{subsec:1NN}

\begin{figure}
	\centering
	\includegraphics[width=1\linewidth]{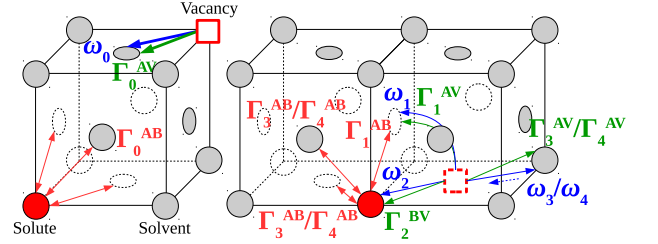}   
	\caption{(Color online) Illustration of all the possible transitions in dilute fcc alloys including 1NN exchanges between atoms and between vacancy and atoms. Red hollow squares designate vacancies V, red filled circles designate solute atoms B, grey filled or hollow circles designate solvent atoms A. }\label{Fig:atomic_jumps_vac_nomencaltures} 
	\vspace{-0.em}
\end{figure}

Here we focus exclusively on the diffusion properties of a dilute binary model alloy A(B): a host matrix of atoms A containing a single solute atom of species B and a single vacancy. The crystallographic structure is chosen to be a fcc crystal. As explained in Sec.\,\ref{subsec:cascade_models}, we consider the vacancy as the only type of PDs. Our purpose is to extend the SCMF theory to include athermal FAR mechanisms. In order to derive analytical transport coefficients, we start with a first shell approximation. This approximation consists in neglecting kinetic coupling and thermodynamic interactions between B and V if the distance between both species is beyond 1NN. FAR-a and FAR-d are restricted to exchanges between 1NN sites only. In such dilute alloy, there are five different atom-vacancy thermal exchange frequencies ($\omega_{i=0,1,2,3,4}$) which we designate after the Lidiard's nomenclature~\cite{Lidiard1955} (see Fig.\,\ref{Fig:atomic_jumps_vac_nomencaltures}), to which we add 4 FAR-a frequencies ($\Gamma_{i=0,1,3,4}^{AB}$) and 5 FAR-d frequencies ($\Gamma_{i=0,1,2,3,4}^{AV}$ and $\Gamma_{2}^{BV}$). {The total B-V exchange frequency is noted $W_2^{BV}$. The total A-V and A-B exchanges conserving the 1NN distance between B and V are respectively noted $W_1^{AV}$ and $W_1^{AB}$. The total A-V and A-B exchanges dissociating the B-V pair are respectively noted $W_3^{AV}$ and $W_3^{AB}$. The total A-V and A-B exchanges associating the B-V pair are respectively noted $W_4^{AV}$ and $W_4^{AB}$, and all the other A-V and A-B exchanges far from the solute atom B are respectively noted $W_0^{AV}$ and $W_0^{AB}$. Here we recall that:
\begin{eqnarray}
    W_i^{AV} && = \omega_i+\Gamma_{i}^{AV} = \omega_i+\Gamma^\text{ad}, \text{ for }i=0,1,3,4 ;\nonumber\\
    W_i^{AB} && = \Gamma_{i}^{AB} = \Gamma^\text{aa}, \text{ for }i=0,1,3,4 ;\nonumber\\
    W_2^{BV} && = \omega_2+\Gamma_{2}^{BV} = \omega_2+\Gamma^\text{ad}.
\end{eqnarray}
}

Within the first shell approximation, two configurational probabilities are considered: $P^{\text{ness}}_1$ for the configuration where B and V are located at 1NN and $P^{\text{ness}}_0$ for the dissociated configuration where B and V are beyond 1NN. The analytical expression of the 1NN-SRO is given by:
\begin{eqnarray}\label{eq:SRO_SCMF}
    \frac{P^{\text{ness}}_1}{P^{\text{ness}}_0} \, &&= \frac{\omega_4+\Gamma^\text{ad}+\Gamma^\text{aa}}{\omega_3+\Gamma^\text{ad}+\Gamma^\text{aa}} \nonumber\\ 
    &&= \frac{\text{exp}\left(E_b/k_B T\right)+(\Gamma^\text{ad}+\Gamma^\text{aa})/\omega_3}{1+(\Gamma^\text{ad}+\Gamma^\text{aa})/\omega_3},
\end{eqnarray}
where $E_b$ is the B-V 1NN binding energy which is deduced from the ratio of thermal frequencies, with $\text{exp}\left(E_b/k_B T\right)=\omega_4/\omega_3$. Note that $P^{\text{ness}}_1/P^{\text{ness}}_0$ is a SRO parameter revealing the binding tendency of B and V at the 1NN. As $\Gamma$ increases, the SRO parameter decreases towards $1$. Note that the decrease in the above-threshold situation is larger than what is expected in the sub-threshold situation, just because two relocation frequencies contribute to the decrease in the above-threshold case. 

The expressions of the phenomenological coefficients $\lambda_{BV}$, $\lambda_{VB}$, $\lambda_{VV}$ and $\lambda_{BB}$ in a dilute binary fcc alloy are given by 
\begin{eqnarray} 
\label{eq:SCMF_1NN_LVB}
\lambda_{VB}= && - \frac{a_0^2}{4}\,{C_{BV}^p} \left[W_2^{BV} - \frac{ \Lambda_4^{B} (\Lambda_3^{V}+\Lambda_4^{V}) }{\Lambda}\right] ,\\ 
\label{eq:SCMF_1NN_LBV}
\lambda_{BV}= && - \frac{a_0^2}{4}\,{C_{BV}^p} \left[W_2^{BV} - \frac{ \Lambda_4^{V} (\Lambda_3^{B}+\Lambda_4^{B}) }{\Lambda}\right] ,\\
\label{eq:SCMF_1NN_LVV}
\lambda_{VV}= && \frac{a_0^2}{4} \left\{{C_V^m W_0^{AV}} + {C_{BV}^p} \left[ W_2^{BV} - \frac{ \Lambda_4^{V} (\Lambda_3^{V}+\Lambda_4^{V}) }{\Lambda}\right]\right\},\nonumber\\
\label{eq:SCMF_1NN_LBB}
\lambda_{BB}= && \frac{a_0^2}{4} \left\{{C_B^m W_0^{AB}} + {C_{BV}^p} \left[ W_2^{BV} - \frac{ \Lambda_4^{B} (\Lambda_3^{B}+\Lambda_4^{B}) }{\Lambda}\right]\right\},\nonumber
\end{eqnarray} 
{where $\Lambda=7W_3 + 2W_1  + 2W_2^{BV}$, $\Lambda_3^{\alpha} = 3 W_3^{A\alpha} - 2 W_1^{A\alpha} -  W_2^{BV}$ and $\Lambda_4^{\alpha} = 3 W_4^{A\alpha}P^\text{ness}_0/P^\text{ness}_1 - 2 W_1^{A\alpha} -  W_2^{BV}$ $(\text{for}~\alpha=B,V)$, with $W_{i}=W_{i}^{AV}+W_{i}^{AB}$ for $i=0,1,3,4$. $C_{BV}^p$ is the concentration of B-V pair at 1NN distance and $C_V^m$ (resp. $C_B^m$) is the concentration of isolated V (resp. B). These concentrations can be deduced from the total concentrations of B and V (resp. $C_B$ and $C_V$) by a low temperature expansion formalism \cite{Sykes1973,Ducastelle1991,Schuler2017}:
\begin{equation}
\left\{
             \begin{array}{lr}
             C_{BV}^p= C_B^0 C_V^0 z^\text{ness} \\
             C_V^m = C_V-C_{BV}^p \\
             C_B^m = C_B-C_{BV}^p,
             \end{array}
\right.
\end{equation} 
with $C_B^0$, $C_V^0$ to be obtained by solving the following system of equations:
\begin{equation}
\left\{
             \begin{array}{lr}
             C_B =  C_B^0 + C_B^0 C_V^0 \left(z^\text{ness}-z_o\right) \\
             C_V =  C_V^0 + C_B^0 C_V^0 \left(z^\text{ness}-z_o\right),
             \end{array}
\right.
\end{equation}}where $z^\text{ness}=12P^\text{ness}_1/P^\text{ness}_0$ is the effective partition function at NESS and $z_0=12$. 

Note that the term $\Lambda_{m=3,4}^V$ (resp. $\Lambda_{m=3,4}^B$) is related to the vacancy (resp. solute atom) mobility since it contains all the vacancy (resp. solute atom) jump mechanisms including A-V (resp. A-B) and B-V (resp. V-B) exchanges. At equilibrium, $\Lambda_{3}^V=\Lambda_{4}^V$ and $\Lambda_{3}^B=\Lambda_{4}^B$ due to the microscopic detailed balance. Hence the two off-diagonal equilibrium coefficients $\lambda_{VB}$ and $\lambda_{BV}$ are equal, according to the Onsager reciprocal relations. In addition, $\lambda_{VV}$ (resp. $\lambda_{BB}$) can be separated into two parts: $C_V^m  W_0^{AV}$ (resp. $C_B^m  W_0^{AB}$) and the rest. The latter represents the exchanges of the B-V pair at 1NN distance while the former represents the hops of the isolated V (resp. B). 

In the case of sub-threshold irradiation for which there is no direct exchange between atoms (i.e. $\Gamma^\text{aa}=0$), the off-diagonal coefficients are equal and, from Eq.\,\eqref{eq:SCMF_1NN_LVB},\,\eqref{eq:SCMF_1NN_LBV} we get:
\begin{equation}\label{eq:LBV_sub}
    \lambda_{BV}= \lambda_{VB}=  -  \frac{a_0^2 C_{BV}^p}{4} \,\frac{W_2^{BV}\left(13W_3^{AV}-2W_1^{AV}\right) }{7  W_3^{AV} + 2 W_1^{AV}  + 2 W_2^{BV}}.
\end{equation}
Although the microscopic detailed balance is broken for the individual exchange frequencies $\omega_3$ and $\Gamma^\text{ad}$, it still holds for the sum of the latter frequencies, that is $W_3=W_3^{AV}=\omega_{3}+\Gamma^\text{ad}$ (see Eq.\,\eqref{eq:SRO_SCMF}). 

By replacing the total frequencies by the corresponding thermally activated jump frequencies, and replacing the dynamical SRO at NESS by the equilibrium SRO, the transport coefficients turn out to be equivalent to the Onsager coefficients $L_{BV}$ of the five-frequency model within the first shell approximation~\cite{Howard1964}.

The variation of $\lambda_{BV}$ with $\Gamma^\text{ad}$ depends on the full set values of the thermal-activated jump frequencies. When $\Gamma^\text{ad}$ is dominant before all $\omega_i$: $\lambda_{BV}\sim -C_B C_V \Gamma^\text{ad}$. Note that if $13\omega_{3}>2\omega_1$ ($L_{BV}<0$), then $\lambda_{BV}$ remains negative whatever the magnitude of the relocation frequencies. Otherwise, a change of sign of $\lambda_{BV}$ can be observed when $\Gamma^\text{ad}\simeq -(13\omega_{3}-2\omega_1)/11$. Therefore, when a solute atom is dragged by a vacancy, FAR-d may change the sign of the solute-vacancy flux coupling and destroy the solute drag effect. In the opposite case, when $L_{BV}$ is negative, FAR-d does not change the sign of the solute-vacancy flux coupling. 

We consider now the case of above-threshold irradiation. Then FAR have two contributions: FAR-a and FAR-d, with $\Gamma^\text{aa}=\Gamma$ and $\Gamma^\text{ad}=\gamma\Gamma$. The off-diagonal terms $\lambda_{BV}$ and $\lambda_{VB}$ are not equal and their difference $\Delta\lambda=\lambda_{VB}-\lambda_{BV}$ is given by:
\begin{equation}
    \Delta\lambda=\frac{3a_0^2 C^p_{BV}}{4}\,\frac{(1-\omega_3/\omega_4)[\omega_a-(1-\gamma)\Gamma]\,\omega_4\,\Gamma}{(\omega_b+11\gamma\Gamma+9\Gamma)\left[\omega_4+(1+\gamma)\Gamma\right]},
\end{equation}
with
\begin{eqnarray}
    \omega_a&&=2\omega_2+2\omega_1-3\omega_3, \nonumber \\
    \omega_b&&=7\omega_3+2\omega_1+2\omega_2.
\end{eqnarray}
Note that $\Delta\lambda=0$ in the two extreme cases when thermal jumps ($\omega$) are dominant (i.e. $\Gamma/\omega\rightarrow 0$) or negligible ($\Gamma/\omega\rightarrow \infty$). The sign of $\Delta\lambda$ is determined by the product $(1-\omega_3/\omega_4)[\omega_a-(1-\gamma)\Gamma]$. If $\gamma=1$, this product involves thermal jump frequencies only. The first parenthesis is directly related to the equilibrium SRO parameter: $(1-\omega_3/\omega_4)$ is positive if the vacancy and the solute atom attract each other and negative otherwise. The higher the thermodynamic attraction, the smaller the ratio $\omega_3/\omega_4$, and the larger the difference $\Delta\lambda$.

\subsection{Extension of the KineCluE code} \label{subsec:KineCluE}

{For a more precise calculation beyond the first shell approximation, we consider each configuration where V and B are located at a distance lower than the kinetic radius $R_k$, as B-V pair configuration. At distances larger than $R_k$, B and V are considered as isolated monomers. Therefore, 3 cluster contributions are included: monomer B, monomer V as well as B-V pair. Note that the calculation under first shell approximation performed at Sec.\,\ref{subsec:1NN} is a particular situation where the kinetic radius is set equal to the 1NN distance. The calculation of the cluster transport coefficients is performed using the KineCluE code~\cite{Schuler2020}. The latter accounts for all the kinetic paths within a pair cluster defined by radius $R_k$.} Note that the kinetic radius can be set well beyond the 1NN distance in KineCluE. This allows us to perform converged calculation of cluster transport coefficients including long-distance FAR as well as long range kinetic correlations. In order to use NESS as reference state, a module is added to the code which calculates the NESS probability distribution by solving Eq.\,\eqref{eq:master_equation}. Besides, the underlying principle of the microscopic detailed balance of the code is replaced by the global detailed balance condition (Eq.\,\eqref{eq:global_detailed_balance}). Detailed descriptions will be published elsewhere. {Models 1, 2 and 3 presented in Section\,\ref{subsec:cascade_models} have been introduced into KineCluE. Note that the cluster radius $R_c$ in Model 3 is set equal to $R_k$ for simplicity.}

\subsection{Comparison between KineCluE results and Monte Carlo simulations} \label{subsec:AKMC_and_validation}
{
As mentioned in the introduction (Sec.\,\ref{sec:introduction}), as soon as one of the microscopic diffusion mechanisms ($W_{\textbf{n}\rightarrow \widetilde{\textbf{n}}}^{AV}$ and $W_{\textbf{n}\rightarrow \widetilde{\textbf{n}}}^{AB}$) does not obey the microscopic detailed balance, we cannot use the Allnatt formulae~\cite{Allnatt1965,Allnatt1982} to extract the phenomenological transport coefficients from atomistic kinetic Monte Carlo (AKMC) simulations. However, for a binary alloy with solute-point defect interactions restricted to 1NN pairwise interactions, we have shown in Sec.\,\ref{subsec:1NN} that detailed balance is fulfilled in the case of sub-threshold irradiation. Therefore, in this specific case, we may rely on the Allnatt formulae to obtain the Onsager matrix of the transport coefficients. As for the thermodynamic properties, we may apply AKMC to study the dynamical short range order characterizing a NESS from an average on the residential time, relying on the ergodic principle. 

We choose here a model alloy with highly attractive vacancy-solute interactions because it emphasizes the effect of FAR on flux coupling. The migration barriers (in eV) are set to 0.95 for $\omega_0$ and $\omega_3$, 0.75 for $\omega_1$ and $\omega_4$, and 0.60 for $\omega_2$. The attempt frequency $\nu$ is chosen to be $10^{14}~\text{s}^{-1}$. As for the model of FAR, we choose Model 1, with $r_m$ equal to the 1NN distance $r_1$. 

The AKMC simulation box is a fcc crystal of 2048 sites. It contains one single solute atom and one vacancy. We apply periodic boundary conditions and use a residence-time algorithm. At each Monte Carlo step, we propose the whole set of the thermal jumps and FAR. We select one exchange from the proposed mechanisms. After every exchange, we compute the residence time increment. From the fluctuations of atomic positions, we compute the transport coefficients. Note that the corresponding off-diagonal coefficients given by the AKMC method are by construction symmetric. As shown in Ref.\,\cite{Gallavotti1996,Lau2007}, they do not correspond to the transport coefficients $\lambda_{BV}$ and $\lambda_{VB}$ whenever one of the diffusion mechanism does not obey the detailed balance.  

As for the KineCluE approach, the kinetic radius $R_k$ of the cluster B-V is set to $4a_0$. 
}
\begin{figure} 
	\centering
	\includegraphics[width=1.0\linewidth]{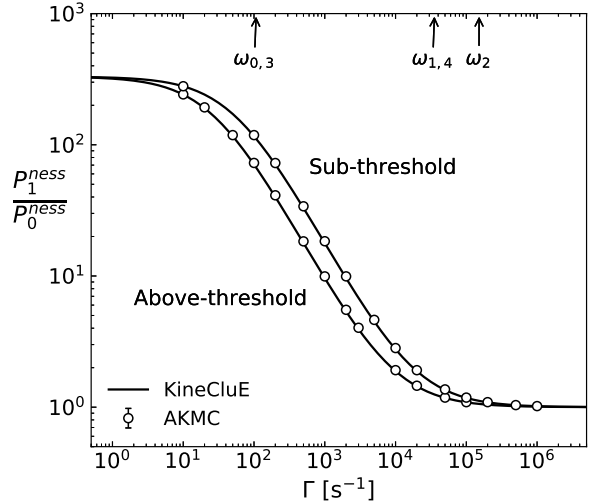}   
	\caption{First nearest neighbor (1NN) short range order as a function of FAR frequency $\Gamma$ from KineCluE and AKMC simulations. Results are obtained for $\omega_4=3.55\times 10^4s^{-1}$ and $\omega_3=1.07\times 10^2s^{-1}$ at $T=400$K. Model 1 is applied.} \label{Fig:SRO_1}
	\vspace{-0.em}
\end{figure}

Fig.\,\ref{Fig:SRO_1} shows the evolution of the dynamical 1NN-SRO under sub- and above-threshold FAR. We obtain an excellent agreement between KineCluE and AKMC simulations on the SRO paramters. As expected, the dynamical SRO decreases with the relocation frequency with a higher rate in the case of an above-threshold irradiation.   

\begin{figure} 
	\centering
	\includegraphics[width=1.0\linewidth]{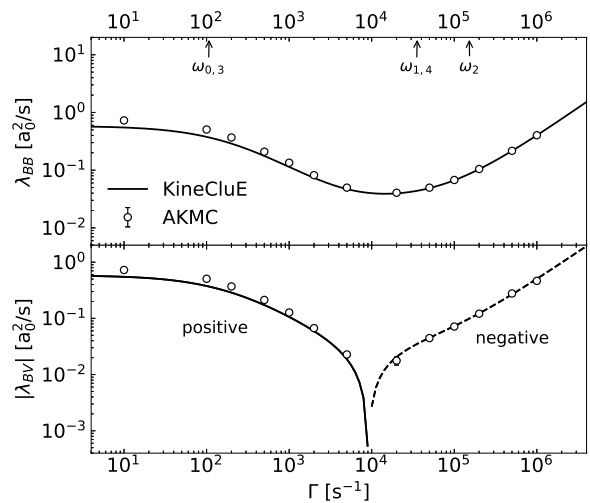}   
	\caption{Solute atom diffusion coefficient and off-diagonal coefficients of transport matrix as a function of the FAR-d frequency $\Gamma$ from KineCluE (solid and dashed lines) and AKMC (unfilled circles) simulations. Results are obtained for $\omega_{0,3}=1.07\times 10^2s^{-1}$, $\omega_{2}=1.52\times 10^5s^{-1}$ and $\omega_{1,4}=3.55\times 10^4s^{-1}$ at $T=400$K. Model 1 is applied, with 1NN FAR-d only.} \label{Fig:bal_sub_DB_LBV_LVB}
	\vspace{-0.em}
\end{figure}

Fig.\,\ref{Fig:bal_sub_DB_LBV_LVB} shows the variation of the transport coefficients with the frequency of FAR-d in the sub-threshold irradiation regime. Both KineCluE and AKMC methods give the same transport coefficients because the microscopic detailed balance holds for the total transition rates. However, when $\Gamma$ is small compared with thermal jump frequencies, we observe a slight discrepancy between the coefficients. Yet the size of the AKMC simulation box is comparable with $R_k$. The discrepancy is due to the difference in the applied boundary conditions between KineCluE and the AKMC method. In KineCluE, configurations of solute and vacancy located at a distance larger than the kinetic radius are not included in the calculation, while the AKMC method relies on periodic boundary conditions. In the latter, atoms or PDs exiting from the simulation box enter back through another side and add a kinetic correlation contribution to the transport coefficients. 

\begin{figure} 
	\centering
	\includegraphics[width=1.0\linewidth]{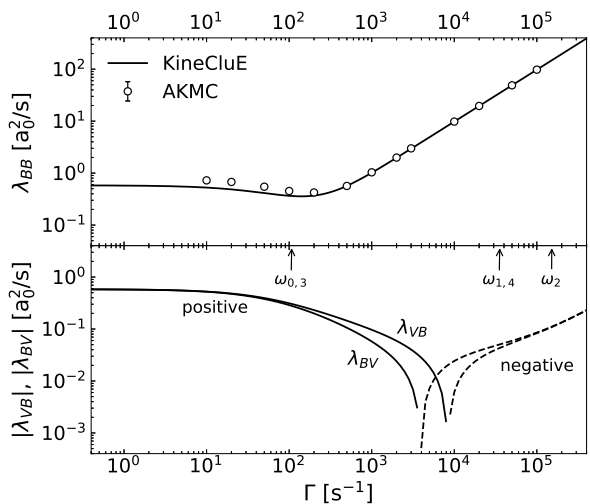}   
	\caption{Solute atom diffusion coefficient and off-diagonal coefficients of transport matrix as a function of the above-threshold relocation frequency $\Gamma$ from KineCluE (solid and dashed lines) and AKMC (unfilled circles) simulations. Results are obtained for $\omega_{0,3}=1.07\times 10^2s^{-1}$, $\omega_{2}=1.52\times 10^5s^{-1}$ and $\omega_{1,4}=3.55\times 10^4s^{-1}$ at $T=400$K. Model 1 is applied, with 1NN FAR-d and FAR-a.} \label{Fig:bal_dir_DB_LBV_LVB}
	\vspace{-0.em}
\end{figure}

In the case of an above-threshold irradiation, we observe in Fig.\,\ref{Fig:bal_dir_DB_LBV_LVB} a similar behaviour of the diagonal transport coefficient $\lambda_{BB}$, whereas the single off-diagonal coefficient measured in AKMC simulations does not correspond any more to the off-diagonal phenomenological transport coefficients obtained by KineCluE.

\section{Results on diffusion properties} \label{sec:results}
Here we focus on the above-threshold irradiation case. We consider a model alloy with relatively high migration barriers. Hence the alloy is potentially sensitive to FAR effects, just because the thermal jump frequencies are small with respect to the relocation frequency deduced from realistic dose rate. The energy interaction between B and V is restricted to a pairwise 1NN interaction. The migration barriers (in eV) are set to 1.10 for $\omega_0$, $\omega_1$ and $\omega_3$, 0.90 for $\omega_4$, and 0.80 for $\omega_2$. The attempt frequency $\nu$ is chosen to be equal to $5\times10^{12}~\text{s}^{-1}$. The three relocation models indicated in the Section~\ref{subsec:cascade_models} are considered. We use KineCluE to calculate the transport coefficients.

The parameter values that we set to estimate the vacancy concentration under irradiation are shown in Table\,\ref{tab:vacancy_concentration_parameter}. Here the mean relocation range $r_m$ and the cut-off distance $L$ for Models 2 and 3 are respectively set to 1NN ($\sqrt{1/2}\,a_0$) and 5NN ($\sqrt{5/2}\,a_0$) distances. The kinetic radius is set to $2a_0$. 
\begin{table}[b]
\caption{\label{tab:vacancy_concentration_parameter}%
List of the parameters needed to estimate the vacancy concentration and their values set in the paper. 
}
\begin{ruledtabular}
\begin{tabular}{ll}
\textrm{Parameter} &
\textrm{Value} \\
\colrule
Lattice parameter $a_0$  &  0.35\,nm\\
Vacancy formation enthalpy $H_{V}^{\text{f}}$ & 1.65\,eV\\
Vacancy formation entropy $S_{V}^{\text{f}}$ & 1.82\,$k_B$\\
Number of relocations per dpa $n_\text{rel}$ & 100\\
Effective sink strength $k^2$ & $10^{15}\,\text{m}^{-2}$\\
\end{tabular}
\end{ruledtabular}
\end{table}

\subsection{Dynamical short range order} \label{subsec:dynamical_SRO}

\begin{figure} 
	\centering
	\includegraphics[width=1.0\linewidth]{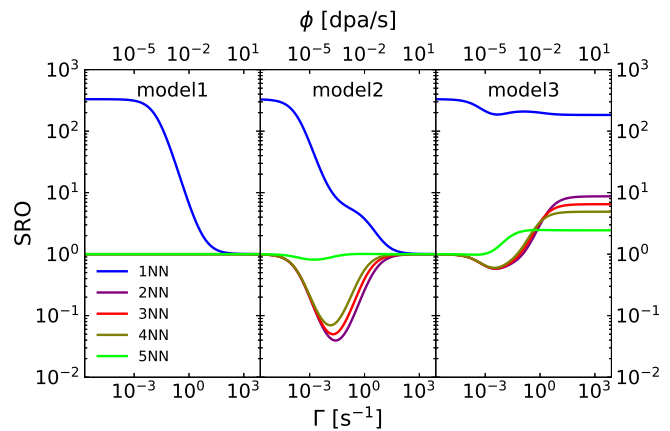}   
	\caption{(Color online) Steady-state short range order as a function of the relocation frequency in the above-threshold radiation regime. Results are obtained by KineCluE for $\omega_{0,1,3}=6.9\times 10^{-2}s^{-1}$, $\omega_2=4.2\times 10^{2}s^{-1}$ and $\omega_4=2.3\times 10^1s^{-1}$ at $T=400$K. The mean relocation range $r_m$ is set 1NN. The cut-off relocation distance and the kinetic radius of the cluster B-V are set to 3$a_0$.} \label{Fig:bal_dir_SRO_models}
	\vspace{-0.em}
\end{figure}
Fig.\,\ref{Fig:bal_dir_SRO_models} shows the profile of dynamical SRO as a function of $\Gamma$ for Models 1, 2 and 3. The probability for B and V at 1NN distance is reduced by FAR leading to an effective B-V interaction smaller than the thermodynamic one. The decrease of 1NN-SRO with the relocation frequency in Model 1 starts when $\Gamma$ is around $10^{-2}\,\text{s}^{-1}$. The decrease starts earlier in Models 2 and 3: respectively around $10^{-4}\,\text{s}^{-1}$ and $10^{-3}\,\text{s}^{-1}$. However, the 1NN-SRO of Model 3 converges towards non-zero value at large $\Gamma$. In Model 1, there is no interaction between B and V beyond the 1NN distance, whatever the relocation frequency. However, in Models 2 and 3, we observe that the effective B-V interaction extends beyond the range of the thermal one (i.e. beyond the 1NN). The effective interaction remains up to 5NN distance when $\Gamma$ is comparable to one of the thermal jump frequencies. This is due to the relatively long range FAR. In the extreme case when $\Gamma$ is dominant before the thermal jump frequencies, the B-V interactions are dropping in Models 1 and 2 whereas in Model 3 the 1NN attraction is slightly decreasing and the 2NN, 3NN, 4NN and 5NN are slightly increasing. The binding tendency of a vacancy around the solute atom is still high ($P_1^\text{ness}/P_0^\text{ness} \simeq 10^2$) due to the introduction of the biased FAR-d with the 1NN atoms of the solute atom in Model 3. 

\subsection{Tracer diffusion coefficient}
In the dilute limit, the tracer diffusion coefficient of solute B is written as
\begin{equation}
    D_B^{*}=\frac{\lambda_{BB}}{C_B}.
\end{equation}
Phenomenological models of diffusion under irradiation systematically rely on the assumption that the thermally activated diffusion and FAR take place in parallel~\cite{Muller1988,Roussel2002}. {The tracer diffusion coefficient is then written as a sum of two diffusion coefficients:
\begin{equation} \label{eq:classical_assumption}
    D_{B,\text{add}}^{*}=D_{B,\text{th}}^{*}C_V^\text{ness}/C_V^\text{eq}+D_{B,\text{far}}^{*},
\end{equation}
where $D_{B,\text{th}}^{*}$ is the thermal diffusion coefficient commonly deduced from diffusion experiments or atomic based diffusion models and $D_{B,\text{far}}^{*}$ is the diffusion coefficient of solute atom B resulting from FAR only.} Note that both coefficients can be calculated by KineCluE. 
Unless one diffusion mechanism is dominant over the other, we expect a non-additive contribution to the solute tracer diffusion coefficient because of the kinetic correlations. In order to quantify the non-additive contribution, we define the parameter:
\begin{equation}
    \Delta D_B = \frac{D_{B,\text{add}}^*-D_B^*}{D_B^*}.
\end{equation}

\begin{figure} 
	\centering
	\includegraphics[width=1.0\linewidth]{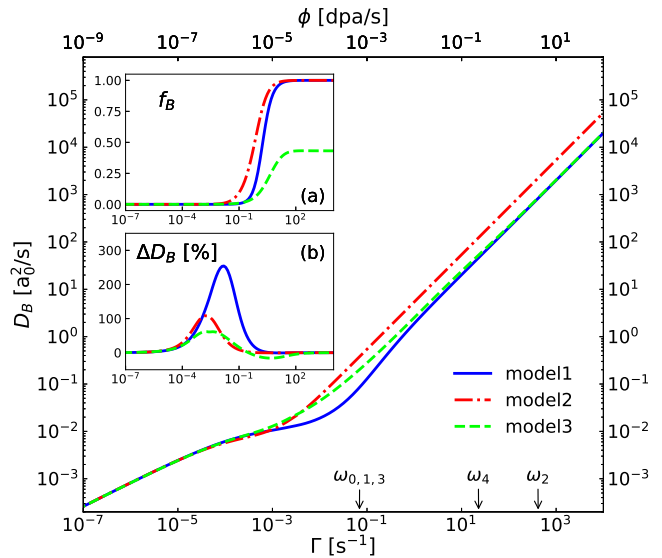}   
	\caption{Solute atom diffusion coefficient as a function of the relocation frequency $\Gamma$ in the above-threshold radiation regime. Results are obtained by KineCluE for $\omega_{0,1,3}=6.9\times 10^{-2}s^{-1}$, $\omega_2=4.2\times 10^{2}s^{-1}$ and $\omega_4=2.3\times 10^1s^{-1}$ at $T=400$K. $C_B$ is set 0.1 at.\%. The mean relocation range $r_m$ is set 1NN. The cut-off relocation distance and the kinetic radius of the cluster B-V are set to 3$a_0$. The insets (a) and (b) show the variations of correlation factor $f_B$ and the difference $\Delta D_B$ with the relocation frequency.} \label{Fig:bal_dir_DB_models}
	\vspace{-0.em}
\end{figure}

Fig \ref{Fig:bal_dir_DB_models} shows the variation of the solute diffusion coefficient with the relocation frequency. We observe that the global tendencies of the diffusion coefficients obtained with the three models are similar. However, the three curves do not have the same asymptote at large $\Gamma$. The largest difference occurs when the correlation factor $f_B$ is increased by FAR. With Models 1 and 2, this factor tends to 1 when $\Gamma$ is dominant over the thermal jump frequencies, meaning that there are no kinetic correlations. However, in Model 3, the correlation factor tends to 0.46. The remaining kinetic correlations are due to the biased FAR-d. Besides, $\Delta D_B$ is high when $\Gamma$ is in the range of the thermal jump frequencies because then, there is a strong competition between the thermal mechanisms and FAR. In this example, $\Delta D_B$ spans from 100\,\% to 300\,\% depending on the relocation model.

\subsection{Flux coupling}

We characterize the flux coupling between solute B and vacancy V by computing the wind factors~\cite{Anthony1969,Anthony1970,Okamoto1979}
\begin{equation}
    \delta_{B\rightarrow V} = \frac{\lambda_{BV}}{\lambda_{VV}}
\end{equation}
and 
\begin{equation}
    \delta_{V\rightarrow B} = \frac{\lambda_{VB}}{\lambda_{BB}}.
\end{equation}
Both wind factors describe the B-V flux coupling related to two different situations. The wind factor $\delta_{B\rightarrow V}$ gives the number of solute atoms following a vacancy under the driving force $\nabla\mu_V$ and the wind factor $\delta_{V\rightarrow B}$ indicates the number of vacancies dragged by a solute atom under the driving force $\nabla\mu_B$. If the wind factors are positive, a drag of B by V (or vice versa) may occur. As shown in Sec.\,\ref{subsec:dynamical_SRO}, the interactions between the solute atom and the vacancy are reduced or even destroyed by FAR. Since the drag effect is highly related to this interaction, we study the effect of the relocation frequency $\Gamma$ on the wind factors.

\begin{figure} 
	\centering
	\includegraphics[width=1.0\linewidth]{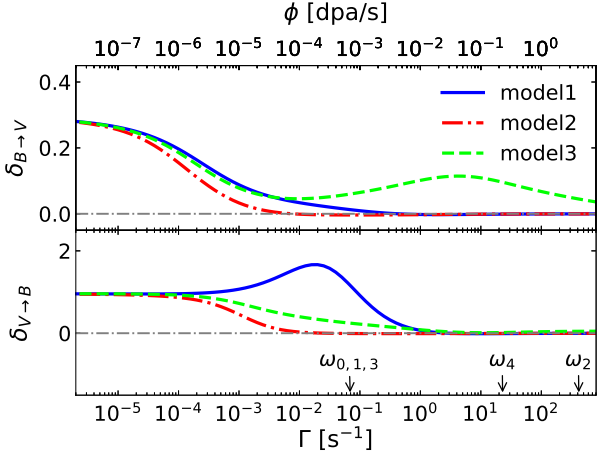}   
	\caption{Drag factors $\delta_{B\rightarrow V}$ and $\delta_{V\rightarrow B}$ as a function of the relocation frequency $\Gamma$ in the above-threshold radiation regime. Results are obtained by KineCluE for $\omega_{0,1,3}=6.9\times 10^{-2}s^{-1}$, $\omega_2=4.2\times 10^{2}s^{-1}$ and $\omega_4=2.3\times 10^1s^{-1}$ at $T=400$K. $C_B$ is set 0.1 at.\%. The mean relocation range $r_m$ is set 1NN. The cut-off relocation distance and the kinetic radius of the cluster B-V are set to 3$a_0$. The dashed lines are eye-guides for $\delta_{B\rightarrow V}=0$ or $\delta_{V\rightarrow B}=0$.} \label{Fig:bal_dir_delta_models}
	\vspace{-0.em}
\end{figure}

Fig.\,\ref{Fig:bal_dir_delta_models} shows the variation of the wind factors with the relocation frequency. Whatever the relocation models, $\delta_{B\rightarrow V}$ and $\delta_{V\rightarrow B}$ globally decrease with $\Gamma$. However, $\delta_{V\rightarrow B}$ of Model 1 has a surprising non-monotonous behaviour: the drag effect is enhanced before being destroyed by FAR. $\delta_{B\rightarrow V}$ of Model 3 has also an atypical behaviour: it slightly increases and tends to a non-zero value at large $\Gamma$, meaning that the solute drag and vacancy drag effects are not totally destroyed. This is because the biased FAR-d maintains a flux coupling between B and V. This persistent flux coupling at high radiation flux should be very sensitive to the details of the relocation mechanism.

\section{Sensitivity study with respect to the model and alloy parameters} \label{sec:sensitivity_relocation_model}
FAR models depend on the values of the mean relocation range $r_m$, the kinetic radius $R_k$ of the cluster B-V and the truncation distance $L$. However, the latter parameter is not a physical parameter. Since the relocation frequency exponentially decreases with the distance between B and V (see Eq.\,\eqref{eq:model_2_distribution}), the value of $L$ does not affect the diffusion properties as long as it is large enough. Therefore, we focus here on the sensitivity of the results to the other two parameters: $r_m$ and $R_k$.

\subsection{Kinetic radius}
\begin{figure} 
	\centering
	\includegraphics[width=1.0\linewidth]{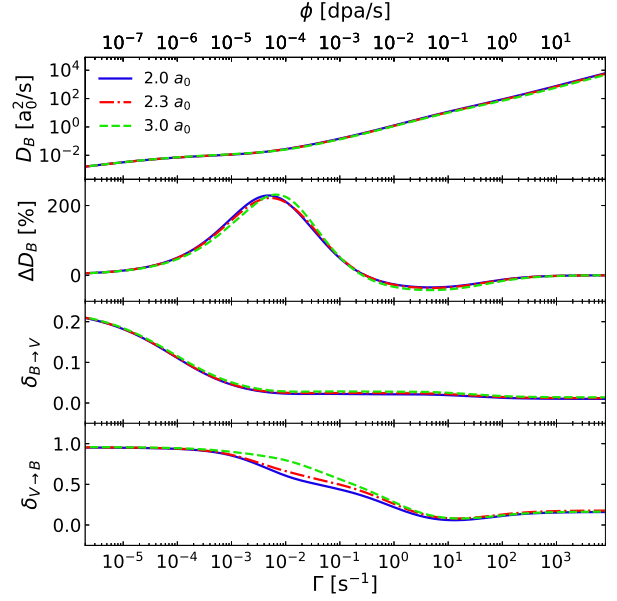}   
	\caption{Diffusion properties as functions of the relocation frequency $\Gamma$ in the above-threshold radiation regime. Results are obtained by KineCluE for $\omega_{0,1,3}=6.9\times 10^{-2}s^{-1}$, $\omega_2=4.2\times 10^{2}s^{-1}$ and $\omega_4=2.3\times 10^1s^{-1}$ at $T=400$K with three different kinetic radius $R_k$ = 2$a_0$, 2.3$a_0$ and 3$a_0$. $C_B$ is set 0.05~at.\%. Model 3 is used as the relocation model. The mean and cut-off relocation distances are respectively set to $(\sqrt{2}/2)a_0$ and 2$a_0$.} \label{Fig:bal_dir_Diffusion_kr}
	\vspace{-0.em}
\end{figure}

In general, the results given by KineCluE code converge with the kinetic radius $R_k$~\cite{Schuler2020}. However, because $R_c=R_k$ in Model 3, the FAR-d models for a monomer vacancy and for a vacancy within the B-V pair are different. In this case, the results obtained with Model 3 may depend on the values of $R_k$. However, Fig.\,\ref{Fig:bal_dir_Diffusion_kr} shows that $D_B^*$, $\Delta D_B$ and $\delta_{B\rightarrow V}$ are not very sensitive to the change of the kinetic radius. Although, the decrease rate of $\delta_{V\rightarrow B}$ with $\Gamma$ is slower with $R_k=3a_0$ than $2a_0$. This is because the vacancy performs biased FAR-d with the 1NN atoms of the solute atom from longer distances.

\subsection{Mean relocation range} \label{subsubsec:rm}
\begin{figure} 
	\centering
	\includegraphics[width=1.0\linewidth]{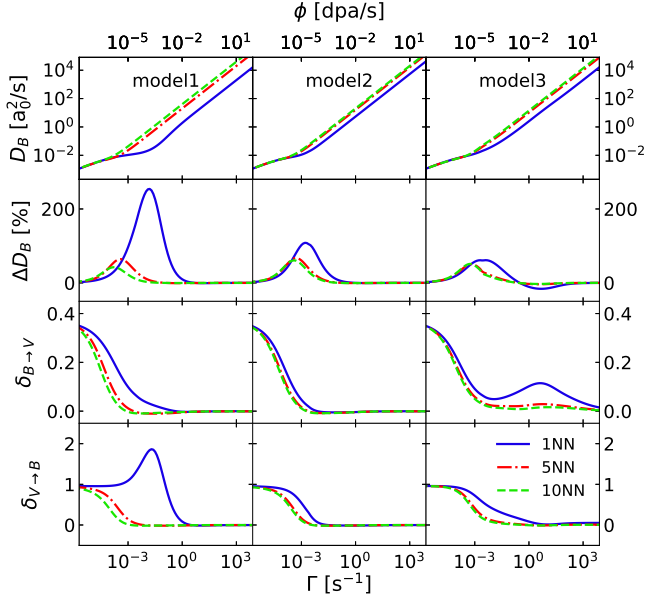}   
	\caption{Diffusion properties as functions of the relocation frequency $\Gamma$ in the above-threshold radiation regime. Results are obtained by KineCluE for $\omega_{0,1,3}=6.9\times 10^{-2}s^{-1}$, $\omega_2=4.2\times 10^{2}s^{-1}$ and $\omega_4=2.3\times 10^1s^{-1}$ at $T=400$K with three different values of $r_m$: 1NN, 5NN and 10NN. $C_B$ is set 0.1 at.\%. The cut-off relocation distance and the kinetic radius of the cluster B-V are set to 3$a_0$. } \label{Fig:bal_dir_Diffusion_rm}
	\vspace{-0.em}
\end{figure}
Fig.\,\ref{Fig:bal_dir_Diffusion_rm} shows the effect of the mean relocation distance $r_m$ on the solute diffusion and flux coupling. First we focus on Model 1. Since the solute mobility is enhanced when increasing the relocation distance, the corresponding solute diffusion coefficient increases with $r_m$. Besides, according to the plot of $\Delta D_B$, the interaction between thermal jumps and FAR decreases with $r_m$. The thermally-activated jump distance and the thermal interaction between B and V are both restricted to 1NN. The larger the relocation distance, the smaller the B-V interaction. Thus B and V are more likely to diffuse as monomers, a kinetic regime where the diffusion properties related to the thermal jumps and FAR become additive. As for the flux coupling, the decreasing rate of $\delta_{B\rightarrow V}$ with $\Gamma$ increases with $r_m$. Thus the solute drag effect is destroyed more easily. Besides, the variation tendency of $\delta_{V\rightarrow B}$ with $\Gamma$ become qualitatively different when $r_m>$ 1NN. The vacancy drag effect is not enhanced when $r_m$ equals to 2NN and 3NN. This may be due to the same reason mentioned before: B and V have many more paths to escape from each other. As for the results obtained with Models 2 and 3, they have similar profiles as the ones in Model 1.

{
\subsection{Atomic mixing rate} \label{subsec:n_bal}
\begin{figure}
	\centering
	\includegraphics[width=1.0\linewidth]{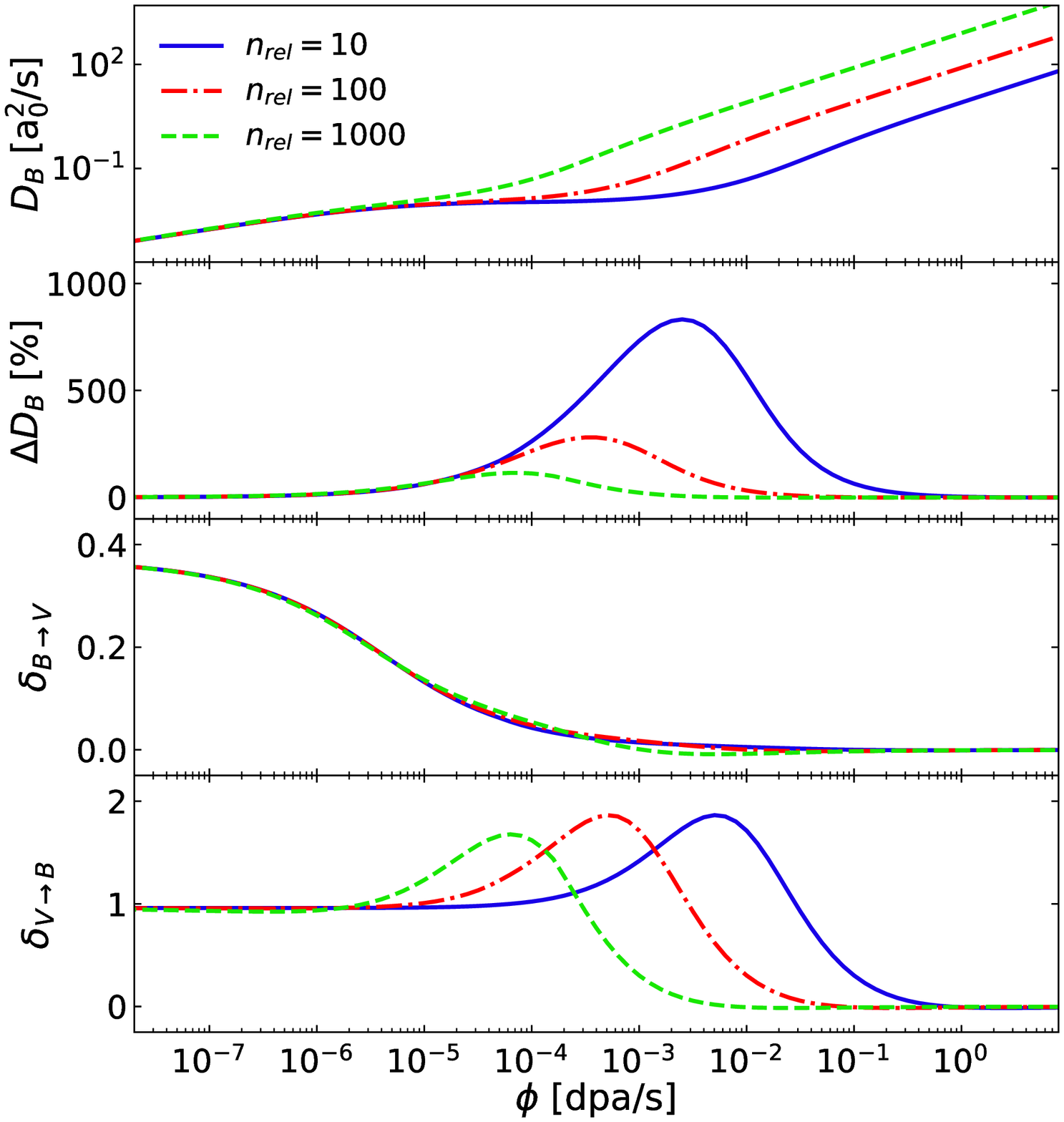}   
	\caption{$D_B$, and wind factors $\delta_{B\rightarrow V}$, $\delta_{V\rightarrow B}$ as a function of relocation frequency $\Gamma$ from KineCluE simulations. Results are obtained for $\omega_{0,1,3}=6.9\times 10^{-2}s^{-1}$, $\omega_2=4.2\times 10^{2}s^{-1}$ and $\omega_4=2.3\times 10^1s^{-1}$ with different values of $n_\text{rel}$ at $T=400$K. $C_B$ is set 0.1 at.\%.} \label{Fig:bal_dir_sensitivity_nbal}
	\vspace{-0.em}
\end{figure}
Since the previous section has shown that the effects of FAR are roughly the same in terms of the global tendency whatever the relocation model and the mean relocation distance, we choose the simplest model, Model 1 with $r_m=r_1$. As stated in Sec.\,\ref{subsec:forced_relocation}, the number of relocations per Frenkel pair created (i.e. $n_\text{rel}$) should be alloy specific due to the thermal effect on heat spike mixing. Fig.\,\ref{Fig:bal_dir_sensitivity_nbal} shows the variation of the diffusion properties in function of radiation dose rate with different values of $n_\text{rel}$. The effect of FAR on the flux coupling and tracer diffusion occurs at a smaller dose rate when $n_\text{rel}$ increases. Moreover, we observe that $\Delta D_B$ decreases with $n_\text{rel}$. These results show the importance of $n_\text{rel}$ in the prediction of a critical dose rate when the effects of FAR on the flux coupling and tracer diffusion is paramount. 

\subsection{FAR-d frequencies}\label{subsec:sensibility_PD_frequency}
\begin{figure}
	\centering
	\includegraphics[width=1.0\linewidth]{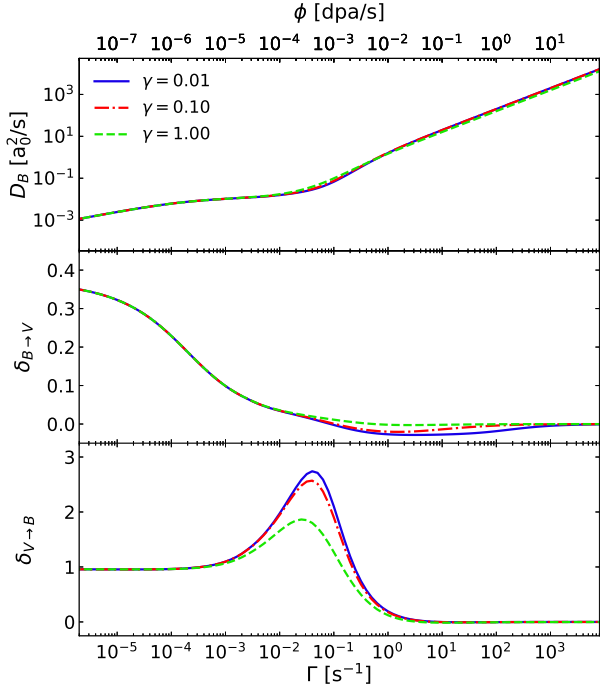}   
	\caption{$D_B$, and wind factors $\delta_{B\rightarrow V}$, $\delta_{V\rightarrow B}$ as a function of relocation frequency $\Gamma$ from KineCluE simulations. Results are obtained for $\omega_{0,1,3}=6.9\times 10^{-2}s^{-1}$, $\omega_2=4.2\times 10^{2}s^{-1}$ and $\omega_4=2.3\times 10^1s^{-1}$ with different values of $\gamma=\Gamma^{ad}/\Gamma^{aa}$ at $T=400$K. $C_B$ is set 0.1 at.\%.} \label{Fig:bal_dir_sensitivity_gamma}
	\vspace{-0.em}
\end{figure}
For the reason stated in Sec.\,\ref{subsec:n_bal}, Model 1 with $r_m=r_1$ is chosen for the sake of simplicity. Note that FAR-a and FAR-d are due to different phenomena in the displacement cascade: FAR-a describes the recoil mixing due to PKA while FAR-d models the lattice site change of PD during the quenching process. There is no guarantee that the frequencies of FAR-a ($\Gamma^{aa}$) and FAR-d ($\Gamma^{ad}$) are equal. Fig.\,\ref{Fig:bal_dir_sensitivity_gamma} shows the plot of $D_B$, $\delta_{B\rightarrow V}$ and $\delta_{V\rightarrow B}$ as a function of relocation frequency $\Gamma^{aa}=\Gamma$ with different ratios $\gamma=\Gamma^{ad}/\Gamma^{aa}$. The global tendencies of the above quantities are not affected by the variation of the ratio $\gamma$. Besides, the tracer diffusion coefficient $D_B$ is not sensitive to the variation of the ratio $\gamma$. However, $\delta_{B\rightarrow V}$ decreases with $\gamma$ while the variation of $\delta_{V\rightarrow B}$ has the opposite tendency. For $\gamma\neq 1$, FAR-a and FAR-d effects on the solute atom and PD diffusion occurs at different dose rate. It respectively happens when $\Gamma^{aa}$ (i.e. $\Gamma$) and $\Gamma^{ad}$ (i.e. $\gamma\Gamma$) are of the same order of magnitude compared with thermal jump frequencies. In brief, the smaller the $\gamma$ value, the larger the difference between the frequencies for FAR-a and FAR-d, and the more important the strength of the flux coupling.
}

\subsection{Thermal jump parameters} \label{subsec:sensitivity_alloy}
\begin{figure}
	\centering
	\includegraphics[width=1.0\linewidth]{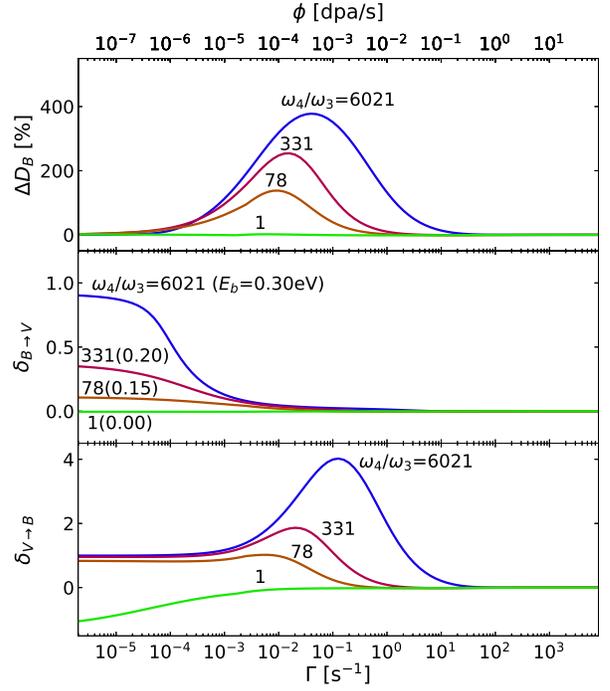}   
	\caption{$\Delta D_B$, and wind factors $\delta_{B\rightarrow V}$, $\delta_{V\rightarrow B}$ as a function of relocation frequency $\Gamma$ from KineCluE simulations. Results are obtained for $\omega_{0,1,3}=6.9\times 10^{-2}s^{-1}$ and $\omega_2=4.2\times 10^{2}s^{-1}$ with different values of $\omega_4$ at $T=400$K. $C_B$ is set 0.1 at.\%.} \label{Fig:bal_dir_sensitivity_binding_energy}
	\vspace{-0.em}
\end{figure}

\begin{figure}
	\centering
	\includegraphics[width=1.0\linewidth]{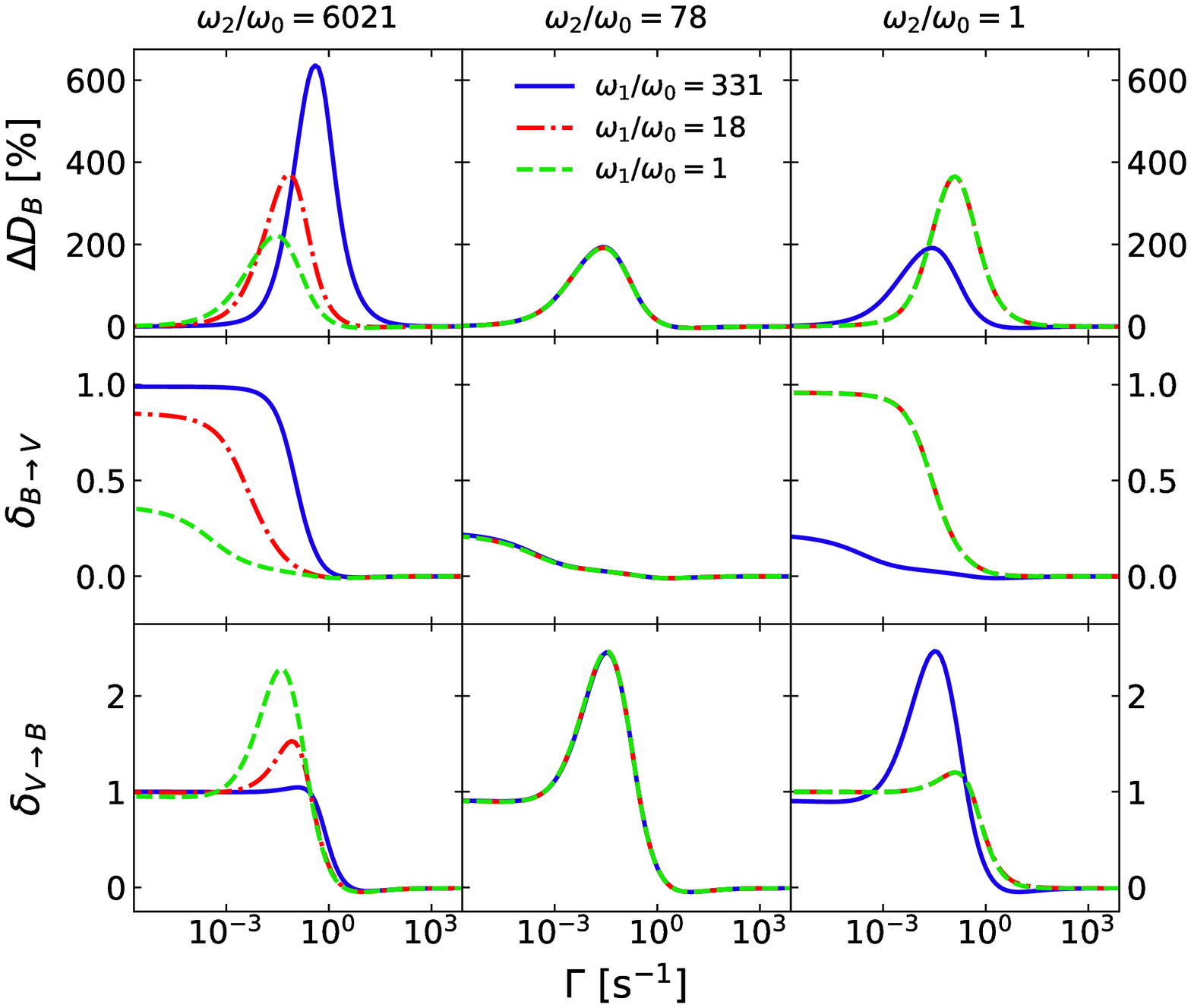}   
	\caption{$\Delta D_B$, and wind factors $\delta_{B\rightarrow V}$, $\delta_{V\rightarrow B}$ as a function of relocation frequency $\Gamma$ from KineCluE simulations. Results are obtained for $\omega_{0,3}=6.9\times 10^{-2}s^{-1}$, $\omega_2=4.2\times 10^{2}s^{-1}$ and $\omega_4=2.3\times 10^1s^{-1}$ with different values of $\omega_1$ and $\omega_2$ at $T=400$K. $C_B$ is set 0.1 at.\%.} \label{Fig:bal_dir_sensitivity_w1_w2}
	\vspace{-0.em}
\end{figure}

The effects of FAR depend on the radiation dose rate and the intrinsic thermal jump frequencies of the alloy. We use KineCluE to perform a sensitive study of the radiation kinetic properties with respect to the thermal jump frequencies. Fig.\,\ref{Fig:bal_dir_sensitivity_binding_energy} shows the variation of $\Delta D_B$ and the wind factors $\delta_{B\rightarrow V}$, $\delta_{V\rightarrow B}$ with respect to $\Gamma$, for various values of $\omega_4$. The values of the other thermal jump frequencies are fixed. Model 1, with $r_m=r_1$, is chosen for the following discussion. The interactions between thermal jumps and FAR are emphasized in this case because the hop distances are both 1NN. The ratio $\omega_4/\omega_3$ directly affects the binding energy $E_b$ between solute atom and vacancy at 1NN. We observe that $\Delta D_B$ and wind factors increase with the binding energy. Besides, the larger the binding energy, the larger the enhancement of the wind factor $\delta_{V\rightarrow B}$ by FAR. This can be explained by noting that the solute atom and vacancy tend to be closer to each other with a larger binding energy. Therefore, the interaction between FAR and thermally activated diffusion of solute atom is more important, leading to a larger difference from what we would except with an additive model, i.e. Eq.\,\eqref{eq:classical_assumption}. Moreover, the binding tendency of the vacancy and the solute atom increases, causing an enhancement of the wind factor $\delta_V$. As well, $\omega_1$ and $\omega_2$ have a non-negligible effect on the profile of $\Delta D_B$ and wind factors in function of $\Gamma$. Here we set $\omega_4$ to its initial value $2.3\times 10^{1}s^{-1}$ and we perform calculations with different values of $\omega_1$ and $\omega_2$. Fig.\,\ref{Fig:bal_dir_sensitivity_w1_w2} shows that if $\omega_2$ is large compared to $\omega_1$ (more than 1 order of magnitude), $\Delta D_B$ and $\delta_{B\rightarrow V}$ increase with $\omega_1$ whereas the enhancement of $\delta_{V\rightarrow B}$ by FAR decreases with $\omega_1$. If the amplitudes of $\omega_2$ and $\omega_1$ are comparable (within 1 order of magnitude), the trends are opposite: $\Delta D_B$ and $\delta_{B\rightarrow V}$ decrease with $\omega_1$ whereas the enhancement of $\delta_{V\rightarrow B}$ by FAR increases with $\omega_1$. However, we observe that if the values of $\omega_1$ and $\omega_2$ are close (within 1 order of magnitude), the variations of $\Delta D_B$ and wind factors with $\Gamma$ are not sensitive to $\omega_1$. 

\section{Discussion and Summary} \label{Sec_Discussions}

Neutron or ion irradiation in metals generates displacement cascades. We present a simplified model of this complex phenomenon by introducing FAR mechanisms, and an average creation rate of PD uniform in time and space. To calculate the energetic and kinetic properties, we write a Master Equation for the evolution of the distribution function which includes both the thermal jumps and FAR. We extend the SCMF theory to solve and compute the SRO parameters and the phenomenological transport coefficients at the NESS reached under irradiation. The main difficulty lies in the loss of the microscopic detailed balance when considering FAR mechanisms. Relying on Model 1 including FAR between 1NN sites only and a first shell approximation of the kinetic correlations, we derive analytical expressions of the phenomenological transport coefficients. We demonstrate that FAR does not produce a simple additive term to the transport coefficients. When the magnitude of the relocation frequency is in the range of the thermal frequencies, FAR interacts with the thermal diffusion mechanism, yielding non-symmetric off-diagonal transport coefficients and a solute tracer diffusion coefficient deviating from a direct sum of the contributions of thermal jumps and FAR. This deviation increases with the solute kinetic correlations. We use the automated code KineCluE to yield a more systematic study of the effect of the range and magnitude of FAR on the kinetic properties, including a sensitivity study with respect to the alloy thermodynamics and the models of relocation and PD production.

Due to the lack of data on the detailed mechanisms of FAR and PD production, we introduce Models 2 and 3 representing two extreme situations, expecting the real situation to be in-between. In Model 2, we assume that FAR is a fully random process while in Model 3, we introduce a biased FAR-d with the 1NN atoms of the solute atoms to reproduce the fact that the vacancy creation within a cascade is partially driven by the vacancy-solute thermodynamic attraction. As a result, part of the vacancy-solute SRO remains which in turns leads to a higher resistance of the vacancy-solute flux coupling to irradiation. Positive flux coupling is the result of strong kinetic correlations, which can be modified by introducing FAR mechanisms. Our sensitivity study shows that the magnitude of the surviving kinetic correlations strongly depends on the details of the biased FAR-d mechanism, while the reduction of correlations and flux coupling due to the randomizing processes are less sensitive to the details of the relocation events unless the distance of FAR is close to the thermodynamic range. A persistent vacancy-solute flux coupling at low temperature and high radiation flux may play an important role on the solute redistribution in irradiated materials. Therefore, the mechanism of PD production with respect to the solute atom spatial distribution within the displacement cascade should be analyzed more precisely.   

Eventually, the effect of the interplay between thermal jumps and FAR on vacancy-solute positive flux coupling is important when the solute-vacancy thermodynamic attraction is large, the magnitude of the thermal jump frequencies compared with the relocation frequency and the range of thermodynamic interactions is close to the relocation distances. As for the tracer diffusion coefficients, their non-additivity property with respect to FAR and thermal jumps follows the same trend as the flux coupling phenomena in systems featuring positive flux coupling but may also arise in case of no positive flux coupling but strong correlated solute migration paths. For instance, the additive expression of Eq.\,\eqref{eq:classical_assumption} reproduces correctly the diffusion coefficient of Au in Al measured under irradiation~\cite{Acker1974}. This is because the vacancy-jump barrier in Al is around 0.58 eV~\cite{Wu2016}, hence the thermal jump frequencies are dominant over the relocation frequencies under realistic experimental conditions. However, we expect a non-negligible effect of FAR in Ni-based alloys because the vacancy-mediated migration barrier in pure Ni is high (around 1.09 eV~\cite{Wu2016}). 

\appendix*
\section{SCMF expressions at NESS}
In this appendix, we introduce in short how to obtain the transport coefficients from the microscopic Master Equation by the standard SCMF formulation~\cite{Nastar2000,Nastar2005}. Note that $\delta P_\textbf{n}(t)$ in Eq.\,\eqref{eq:master_equation_microscopic_2} is a corrective term representing the modification of the effective distribution function $P^{\text{ness}}_\textbf{n}$ due to the presence of an applied driving force. It is written as
\begin{equation} \label{eq:non-steady-state distribution}
    \delta P_\textbf{n}(t) = \text{exp}\left[\beta\left({\delta\Omega(t)+\sum_{\alpha,i}\delta\mu_i^{\alpha}(t)n_i^{\alpha}-h(t)}\right)\right],
\end{equation}
where $\beta=1/k_BT$, $\delta\Omega$ is the normalization factor, $\delta\mu_i^{\alpha}$ is the deviation from the stationary-state chemical potential on site $i$ of the chemical species $\alpha$ compared to the bulk atom, and $h$ is the time-dependent effective Hamiltonian restricted to the pair interaction written as 
\begin{eqnarray}
    h(t)=&&\, \frac{1}{2}\sum_{\alpha,\gamma,i\neq j}\nu_{ij}^{\alpha\gamma}(t)n_i^{\alpha}n_j^{\gamma},
\end{eqnarray}
where $\nu_{ij}^{\alpha\beta}(t)$ is the time-dependent effective pair interactions. The latter can be determined by solving the kinetic equations deduced from the Master Equation. Here, Eq.\,\eqref{eq:non-steady-state distribution} is linearized with respect to the terms $\beta\delta\mu_i^{\alpha}$ and $\beta h$ because we are close to NESS:
\begin{eqnarray}
        \delta P_\textbf{n}(t) = && \, 1+\beta\delta\Omega(t)+\beta\sum_{\alpha,i}\delta\mu_i^{\alpha}(t)n_i^{\alpha} \nonumber\\
        &&-\,\frac{1}{2}\,\beta\sum_{\alpha,\gamma,i\neq j}\nu_{ij}^{\alpha\gamma}(t)n_i^{\alpha}n_j^{\gamma}.
\end{eqnarray}
\begin{widetext}
Starting from the Master Equation Eq.\,\eqref{eq:master_equation_microscopic_2}, the time derivative of the ensemble average can be given by
    \begin{eqnarray}
        \frac{\text{d}}{\text{d}t}\langle n_i^{\alpha}n_j^{\beta}\cdots\rangle = && \, \beta\sum_{\textbf{n},\widetilde{\textbf{n}}}n_i^{\alpha}n_j^{\beta}\cdots P_{\widetilde{\textbf{n}}}^\text{ness} W_{\widetilde{\textbf{n}}\rightarrow \textbf{n}} \left[\sum_{\alpha,i}\delta\mu_{i}^{\alpha}\left(\widetilde{n}_i^{\alpha} - {n}_i^{\alpha} \right) - \,\frac{1}{2}\,\sum_{\alpha,\beta,i\neq j}\nu_{ij}^{\alpha\beta}(t)\left(\widetilde{n}_i^{\alpha}\widetilde{n}_j^{\beta} - n_i^{\alpha}n_j^{\beta}\right) \right],
    \end{eqnarray}
where $\widetilde{n}_i^{\alpha}$ is the occupation number of the configuration $\widetilde{\textbf{n}}$. By denoting $\langle\cdot\rangle^{\text{ness}}$ the ensemble average over the distribution function at NESS (i.e. $P_\textbf{n}^\text{ness}$), the derivative of the one-point average $\langle n_i^{\alpha}\rangle$ can be given by

    \begin{eqnarray}
        \frac{\text{d}}{\text{d}t}\langle n_i^{\alpha}\rangle = && \, \beta \sum_{s\neq i}\sum_\gamma\left< \widetilde{n}_s^{\alpha}\widetilde{n}_i^{\gamma} W_{si}^{\alpha\gamma} \left[ \left( \delta\mu_s^{\alpha\gamma}-\delta\mu_i^{\alpha\gamma}\right) + 2\nu_{si}^{\alpha\gamma} + \,\frac{1}{2}\sum_{\delta,k\neq i\neq s}\left( \nu_{sk}^{\alpha\delta} - \nu_{ik}^{\alpha\delta} + \nu_{ik}^{\gamma\delta} - \nu_{sk}^{\gamma\delta} \right) \right] \right>^\text{ness}, \\
        \frac{\text{d}}{\text{d}t}\langle n_i^{\alpha}n_j^{\gamma}\rangle = && \, \beta \sum_{s\neq i}\sum_\delta\left< \widetilde{n}_s^{\alpha}\widetilde{n}_j^{\gamma}\widetilde{n}_i^{\delta} W_{si}^{\alpha\delta} \left[ \left( \delta\mu_s^{\alpha\delta}-\delta\mu_i^{\alpha\delta}\right) + 2\nu_{si}^{\alpha\delta} + \,\frac{1}{2}\sum_{\epsilon,k\neq i\neq s}\left( \nu_{sk}^{\alpha\epsilon} - \nu_{ik}^{\alpha\epsilon} + \nu_{ik}^{\delta\epsilon} - \nu_{sk}^{\delta\epsilon} \right) \right] \right>^\text{ness} \nonumber \\
        && +  \, \beta \sum_{s\neq i}\sum_\delta\left< \widetilde{n}_i^{\alpha}\widetilde{n}_s^{\gamma}\widetilde{n}_j^{\delta} W_{sj}^{\gamma\delta} \left[ \left( \delta\mu_s^{\gamma\delta}-\delta\mu_j^{\gamma\delta}\right) + 2\nu_{sj}^{\gamma\delta} + \,\frac{1}{2}\sum_{\epsilon,k\neq j\neq s}\left( \nu_{sk}^{\gamma\epsilon} - \nu_{jk}^{\gamma\epsilon} + \nu_{jk}^{\delta\epsilon} - \nu_{sk}^{\delta\epsilon} \right) \right] \right>^\text{ness} \nonumber \\
        && + \, \beta \left< \widetilde{n}_j^{\alpha}\widetilde{n}_i^{\gamma} W_{ji}^{\alpha\gamma} \left[ \left( \delta\mu_j^{\alpha\gamma}-\delta\mu_i^{\alpha\gamma}\right) + 2\nu_{ji}^{\alpha\gamma} + \,\frac{1}{2}\sum_{\delta,k\neq i\neq j}\left( \nu_{jk}^{\alpha\delta} - \nu_{ik}^{\alpha\delta} + \nu_{ik}^{\gamma\delta} - \nu_{jk}^{\gamma\delta} \right) \right] \right>^\text{ness},
    \end{eqnarray}
where $\delta\mu_i^{\alpha\gamma} = \delta\mu_i^{\alpha} - \delta\mu_i^{\gamma}$. By applying the continuity equation to the kinetic equation of the one-point average written as:
\begin{equation}
    \frac{\text{d}}{\text{d}t}\langle n_i^{\alpha}\rangle = - \sum_{s\neq i}J^{\alpha}_{i\rightarrow s},
\end{equation}
we can deduce the expression of the flux of chemical species. The fluxes of solute atom and vacancy under first shell approximation are recognized to be:
    \begin{eqnarray}
        &&J^V_{i\rightarrow s}=- \beta\left<  \widetilde{n}_s^V \widetilde{n}_i^A W_{si}^{VA}\left[ \vec{\nabla}\mu^{VA}\cdot\vec{is} + \frac{1}{2}\sum_{k\neq i\neq s}n_k^B \left( \hat{\nu}_{ks}^{BV} - \hat{\nu}_{ki}^{BV} \right) \right]  +   \widetilde{n}_s^V \widetilde{n}_i^B W_{si}^{VB}\left( \vec{\nabla}\mu^{VB}\cdot\vec{is} + 2 \hat{\nu}_{is}^{BV} \right) \right>^\text{ness}, \\
        &&J^B_{i\rightarrow s}=- \beta\left<  \widetilde{n}_s^B \widetilde{n}_i^A W_{si}^{BA}\left[ \vec{\nabla}\mu^{BA}\cdot\vec{is} + \frac{1}{2}\sum_{k\neq i\neq s}n_k^V \left( \hat{\nu}_{sk}^{BV} - \hat{\nu}_{ik}^{BV} \right) \right]  +   \widetilde{n}_s^B \widetilde{n}_i^V W_{si}^{BV}\left( \vec{\nabla}\mu^{BV}\cdot\vec{is} + 2 \hat{\nu}_{si}^{BV} \right) \right>^\text{ness},
    \end{eqnarray}
where $\hat{\nu}_{ij}^{BV}=\nu_{ij}^{BV}-\nu_{ij}^{BA}-\nu_{ij}^{AV}$.
\end{widetext}
Note that under the first shell approximation, the effective interactions are restricted to the pair B-V at 1NN. The term $\hat{\nu}^{BV}$ can be estimated from the stationary condition of the kinetic equation of the two-point average $\left< n_i^{B}n_j^{V} \right>$. As a result, $\hat{\nu}^{BV}$ can be expressed as a function of the chemical potential gradient. Therefore, the atomic fluxes of solute atom and vacancy are also functions of ${\nabla}\mu^{VA}$ and ${\nabla}\mu^{BA}$, allowing us to identify the transport coefficients and the expressions under first shell approximation are given in Eq.\,\eqref{eq:SCMF_1NN_LVB}--\eqref{eq:SCMF_1NN_LBB}. 

\bibliographystyle{apsrev4-2}
\bibliography{MyRef}

\end{document}